\documentclass[journal]{IEEEtran}
\usepackage{cite}

\usepackage{amssymb}
\usepackage{xcolor}
\ifCLASSINFOpdf
   \usepackage[pdftex]{graphicx}
\else
\fi
\usepackage[fleqn]{amsmath}
\usepackage{algorithmic}

\usepackage{multirow}
\usepackage{array}

\usepackage{stfloats}

\usepackage{nomencl}
\makenomenclature
\begin{document}
%
\title{Time Domain Simulation of DFIG-Based Wind Power System using Differential Transform Method}
%
%
%

\author{Pradeep~Singh,~\IEEEmembership{Member,~IEEE,}
        Upasana~Buragohain, 
         and~Nilanjan~Senroy,~\IEEEmembership{Senior~Member,~IEEE}
\thanks{P. Singh, U. Buragohain and N. Senroy are with the Department
of Electrical Engineering, Indian Institute of Technology, Delhi, 110016 INDIA e-mails: psn121988@gmail.com, upasana.buragohain@ee.iitd.ac.in,  nsenroy@ee.iitd.ac.in.}
}

%
%

\markboth{Journal of \LaTeX\ Class Files,~Vol.~14, No.~8, August~2015}%
{Shell \MakeLowercase{\textit{et al.}}: Bare Demo of IEEEtran.cls for IEEE Journals}
%



\maketitle

\begin{abstract}
This paper proposes a new non-iterative time-domain simulation approach using Differential Transform Method (DTM) to solve the set of non-linear Differential-Algebraic Equations (DAEs) involved in a DFIG-based wind power system. The DTM is an analytical as well as numerical approach applied to solve high dimensional non-linear dynamical systems and the solution can be expressed in the form of a series. In this approach, there is no need to compute higher-order derivatives as DAEs are converted into a set of linear equations after applying transformation rules so that the power series coefficients can be computed directly. The transformation rules are used to transform power system models of various devices, such as induction generator, wind turbine, rotor and grid side converter, which includes trigonometric, square root, exponential functions \textit{etc}. Further, to increase the interval of convergence for the series solutions, the multi-step DTM (MsDTM) approach is used. The numerical performance of the proposed approach is compared with the traditional numerical RK-4 method to demonstrate the potential of the proposed approach in solving power system non-linear DAEs.
\end{abstract}


\IEEEpeerreviewmaketitle

\mbox{}
\nomenclature{$H_{g},H_{t}$}{Generator and turbine inertia constant (\textit{sec.})}
\nomenclature{$\omega_{r},\omega_{t}$}{Generator and wind turbine angular speed (\textit{p.u.})}
\nomenclature{$\omega_{el},\omega_{e},\omega_{s}$}{Electrical base, electrical and synchronous speed (\textit{rad./sec.})}
\nomenclature{$i_{ds},i_{qs}$}{Stator $d$ and $q$-axis current (\textit{p.u.})}
\nomenclature{$i_{dg},i_{qg}$}{$d$ and $q$ axis components of grid side converter current (\textit{p.u.})}
\nomenclature{$i_{dr},i_{qr}$}{$d$ and $q$ axis components of rotor side converter current (\textit{p.u.})}
\nomenclature{$e_{ds}^{'},e_{qs}^{'}$}{$d$ and $q$ axis components of the equivalent voltage source behind transient impedance (\textit{p.u.})}
\nomenclature{$v_{ds},v_{qs}$}{Stator $d$ and $q$ axis voltages (\textit{p.u.})}
\nomenclature{$v_{dr},v_{qr}$}{Rotor $d$ and $q$ axis voltages (\textit{p.u.})}
\nomenclature{$v_{w},v_{wB}$}{Wind and base wind speed (\textit{m/sec.})}
\nomenclature{$K_{Pi},K_{Ii}$}{Proportional and integral gain constant of the $i^{th}$ PI controller}
\nomenclature{$P_{r},P_{g},P_{dc}$}{Active power at RSC, GSC and dc-link (\textit{p.u.})}
\nomenclature{$v_{dg},v_{qg}$}{$d$ and $q$ axis components of grid side converter voltage (\textit{p.u.})}
\nomenclature{$Q_{g},Q_{gref}$}{GSC reactive power and reactive power set-point (\textit{p.u.})}
\nomenclature{$L_{ss},L_{rr},L_{m}$}{Stator, rotor and mutual inductance (\textit{p.u.})}
\nomenclature{$\lambda, \beta$}{Tip speed ratio and blade pitch angle}
\nomenclature{$R_{s}, R_{r}$}{Stator and rotor resistance (\textit{p.u.})}
\nomenclature{$v_{dc},v_{\infty}$}{Capacitor voltage and infinite bus voltage (\textit{p.u.})}
\nomenclature{$T_{m},T_{e},T_{sh}$}{Wind, electromagnetic and shaft torque (\textit{p.u.})}
\nomenclature{$S_{grid},v_{dcref}$}{Complex power delivered to the grid and reference value for the dc voltage (\textit{p.u.})}
\nomenclature{$C_{p}(\lambda,\beta)$}{Wind turbine performance coefficient}
\nomenclature{$\theta_{tw},P_{ref}$}{Shaft twist angle (\textit{rad.}) and active power set point (\textit{p.u.}), which tracks the maximum power point}
\nomenclature{$K_{sh},C_{sh}$}{Drive train shaft stiffness (\textit{p.u./el.rad.}) and damping coefficient (\textit{p.u.sec./el.rad.})}


\printnomenclature

\section{Introduction}
\IEEEPARstart{T}{he} integration of renewable energy sources into the main grid is continuously increasing to reduce the carbon footprints, global warming, dependency on the fossil fuels and unsustainability of conventional energy sources \cite{salman,akhmatov2003analysis}. The global emphasis on the use of clean and green energy have led to increase in initiatives to harness the energy from Wind Energy Conversion System (WECS). Among various state-of-the-art wind generation technologies, the doubly-feed induction generator (DFIG) is most prevalent, because it allows bidirectional power flow, operation on wide range of rotor speed, reduction in the mechanical stress and offers higher energy harness capability \cite{salman,akhmatov2003analysis,hu2015modeling,ymishra,fmei,ghosh2013electromechanical}. These benefits are possible as the DFIG based wind turbine is equipped with the power electronics converters, which consist of a Grid Side Converter (GSC) and a Rotor Side Converter (RSC) in rotor circuit. The controlling capability of these converters permits independent control of both active and reactive powers injected to the grid. However, as the penetration of DFIGs increases, the number of states in the WECS mathematical model also increase, thereby, further increasing the complexity of the power system dynamic model. Moreover, the penetration of WECS also effects the overall stability of the system. Therefore, a detailed investigation of both challenges is a pressing concern for power system researchers and engineers.

For this purpose, the various dynamic models and control algorithms for the DFIG controllers have been developed in the literature \cite{salman,ymishra}. The dynamic performance of the DFIG system has been accessed using traditional numerical integration methods such as Modified Euler method, Runge-Kutta method, Gear method, Trapezoidal method \textit{etc.} with enough small step size to achieve desired accuracy and numerical stability requirements \cite{hu2015modeling,ymishra,fmei,ghosh2013electromechanical}. To suggest corrective measures for any insecure contingency before its occurrence, it is preferred that the dynamic simulation must be transitioned from offline to real time operation. Therefore, more advanced and powerful simulation tools are required by the system operator to perform time domain simulation in real time. In the literature, various techniques such as: a) model reduction \textit{e.g.} selective modal analysis \cite{pulgar2011reduced}, balanced truncation \cite{ghosh2013balanced}, singular perturbation analysis \cite{ni2016model} and coherency-based model reduction technique \cite{chow2020power,osipov2018adaptive}; b) parallel computing \textit{e.g.} the multi-decomposition approach \cite{zadkhast2014multi}, the waveform relaxation method \cite{liu2015two}, the instantaneous relaxation method \cite{Jalili}, and the practical parallel implementation techniques \cite{diao2016parallelizing,konstantelos2016implementation}; c) semi-analytic methods \cite{wang, duan2016power, gurrala, pade} have been suggested to improve the time complexity of dynamic simulation. In model reduction approach, the computational cost is reduced by simplifying the DAEs but the accuracy is decreased. In parallel computing approach, the computational cost is reduced by allocating the computing task to the multiple computer cores, but it requires integration and small time-step to maintain the accuracy. In semi-analytic approach, firstly the semi-analytic solutions are obtained offline and then analytic solutions of DEs are obtained online, but the numerical values of algebraic variables are still determined using numerical iterative method. 

Therefore, a non-iterative approach known as Differential Transform Method (DTM) is developed in the literature to mitigate the aforesaid problems. The Differential Transform (DT) is a mathematical approach by which an approximate solution of a set of linear or non-linear differential, differential-algebraic and partial differential equations can be obtained \cite{chen1999solving, fatoorehchi2012computation, kurnaz2005n,rashidi2009modified, karakocc2009solutions,ev1982differential, zhou1986differential, liu2019power, hassan2008application, ghafarian2016free,xie2016effective, erturk2012multi, odibat2008generalized}. Moreover, DTM is also used to solve DAE model of the power systems and proved to be an adequate tool for the dynamic simulation \cite{liu2019power, xu2020fast, liu2019solving}. The absence of DTM based time-domain simulation approach to assess the dynamic performance of DFIG in the literature, motivates to introduce the DTs of the DFIG. Firstly, in this work the DTs of all differential and algebraic equations are derived and then a non-iterative algorithm is used to solve state and algebraic variables by power series in time. 

The rest of the paper is organized as follows: In section \ref{basicphi}, the basic philosophy and the transformation rules of DT theory have been discussed. Section \ref{proposedDTs} presents the DTs of the DFIG systems. The profiling of the numerical performance of the proposed approach has been addressed in the section \ref{resultsandcon}. Finally, the conclusions of this research work are drawn in section \ref{conclusions}.
 
\section{Basic Philosophy and Transformation Rules of Differential Transform (DT)} \label{basicphi}
The set of non-linear differential algebraic equations (DAEs) arises very frequently to represent the mathematical models of engineering and science problems. In general, the analytic or numerical solution of these problems are difficult to achieve. However, the set of non-linear DAEs can be transformed into the recurrence relations by means of DT which leads to the set of algebraic equations and their approximate solution can be expressed as a finite power series in time \cite{ev1982differential, zhou1986differential, hassan2008application, ghafarian2016free,xie2016effective, erturk2012multi, odibat2008generalized}. The coefficients of this power series can be easily evaluated by the DT without calculating the higher-order derivatives of DEs \cite{xie2016effective, erturk2012multi, odibat2008generalized}.  

\textit{Definition:} If a function $y(t)$ is analytic in a domain $D$ and $t=t_{0}$ is any point in $D$ then the function $y(t)$ can be represented by a power series whose center is located at $t_{0}$. The differential transformation of the $k^{th}$ derivative of a function $y(t)$ is defined as follows:
\begin{small}
\begin{equation}\label{DT}
 Y[k] = \frac{1}{k!} \Bigg [ \frac{d^{k}y(t)}{dt^{k}} \Bigg ]_{t=t_{0}}; \forall t \in D
\end{equation}
\end{small}
where, $y(t)$ is the original function, $Y[k]$ is the transformed function, and $k \in \mathbb{N}$ is the order. The inverse DT of $Y[k]$ is defined by (\ref{IDT}) as follows:
\begin{small}
\begin{equation}\label{IDT}
y(t) = \sum_{k=0}^{\infty}Y[k](t-t_{0})^{k}; \forall t \in D    
\end{equation}
\end{small}
From (\ref{DT}) and (\ref{IDT}), it can be observed that the concept of DT is derived from the Taylor series expansion. The approximate solution of $y(t)$ can be expressed by a finite series (\textit{i.e.} upto $NL$); if $\sum_{k=NL+1}^{\infty}Y[k](t-t_{0})^{k}$ is assumed to be negligible. The approximate solution obtained by DT is valid in the neighborhood of a fixed point $t_{0}$ \textit{i.e.} the series solution doesn't converge for the larger domain. In order to increase the convergence region, MsDTM approach is used \cite{erturk2012multi}. 

The approximate solution of the given initial value problem using DT (using (\ref{IDT})) over the interval $[t_{0}, T]$ can be expressed by a finite series as follows:
\begin{small}
\begin{equation}
y(t) = \sum_{k=0}^{NL}Y[k](t-t_{0})^{k};  t \in [t_{0}, T]    
\end{equation}
\end{small}

Let the interval $[t_{0}, T]$ is divided into $n$ sub-intervals $[t_{i-1}, t_{i}]$, $i=1,2,....,n$ of equal step-size $h=(T-t_{0})/n$. Now, the MsDTM can be applied as follows:\\
1) Firstly, apply the DTM to given differential equation over the interval $[t_{0}, t_{1}]$ and using the initial condition $y(t_{0})=c$, then the approximate solution denoted by $y_{1}(t)$ can be obtained as follows:
\begin{small}
\begin{equation}
 y_{1}(t) = \sum_{k=0}^{NL}Y_{1}[k](t-t_{0})^{k},  t \in [t_{0}, t_{1}]     
\end{equation}
\end{small}
2) For $i\geq 2$, we will use the initial conditions $y_{i}(t_{i-1})=y_{i-1}(t_{i-1})$ at each sub-interval $[t_{i-1},t_{i}]$ and apply the DTM to the given differential equation over the interval $[t_{i-1},t_{i}]$. This process is repeated and generate a sequence of approximate solutions $y_{i}(t)$, $i=1,2,....,n$ for the solution of $y(t)$. Finally, the MsDTM assumes the following solutions:
\begin{small}
\begin{equation}
y(t) = 
\begin{cases}
  y_{1}(t), & t \in [t_{0}, t_{1}] \\
  y_{2}(t), & t \in [t_{1}, t_{2}] \\
  . &  \\
    . &  \\
  y_{n}(t), & t \in [t_{n-1}, t_{n}] 
\end{cases}
\end{equation}
\end{small}
The transformation rules of DT are given in \eqref{DTrules} and for their detailed proof see \cite{erturk2012multi, liu2019power}. In \eqref{DTrules}, the $x(t)$, $y(t)$, $z(t)$, $w(t)$, $\phi(t)$, $\psi(t)$, and $\theta(t)$ are the original functions and $X[k]$, $Y[k]$, $Z[k]$, $W[k]$, $\Phi[k]$, $\Psi[k]$, and $\Theta[k]$ are their DTs, respectively. The symbol $\varrho$ represents the Kronecker delta function; and $c$ and $d$ are constants.
\begin{small}
\begin{equation}
\resizebox{0.45\textwidth}{!}
     {%
$\left.\begin{array}{r@{\;}l}
y(t)|&_{t=0} = y(0) \Rightarrow Y[0] = y(0) \\
y(t) = &  cx(t) \pm dz(t) \Rightarrow Y[k] = cX[k] \pm dZ[k] \\
y(t) = & x(t)z(t) \Rightarrow Y[k] = \displaystyle \sum_{m=0}^{k} X[m]Z[k-m] \\
y(t) = & x(t)z(t)w(t)  \\  \Rightarrow & Y[k] =  \displaystyle \sum_{m=0}^{m_{1}} \sum_{m=m_{1}}^{k} X[m]Z[m_{1}-m]W[k-m_{1}] \\
y(t) = & \frac{x(t)}{z(t)}  \Rightarrow Y[k] = \frac{1}{Z[0]}\bigg[ x[k] - \displaystyle \sum_{m=0}^{k-1}Y[m]Z[k-m]\bigg]\\
y(t) = & t^{n} \Rightarrow Y[k] = \varrho[k-n] = \begin{cases} 1, & k=n \\
																 0, & k\neq n
																 \end{cases} \\
y(t)    = & \frac{d^{n}x}{dt^{n}} \Rightarrow Y[k] = k(k+1)....(k+n)X[k+n] \\
\phi(t) = & sin\theta(t)  \Rightarrow \Phi[k] = \displaystyle \sum_{m=0}^{k-1}\bigg(\frac{k-m}{k}\bigg)\Psi[m]\Theta[k-m] \\
\psi(t) = & cos\theta(t)  \Rightarrow \Psi[k] = \displaystyle - \sum_{m=0}^{k-1}\bigg(\frac{k-m}{k}\bigg)\Phi[m]\Theta[k-m] \\
y(t)  = & e^{x(t)} \Rightarrow Y[k] = \displaystyle \sum_{m=0}^{k-1} \bigg(\frac{k-m}{k}\bigg) Y[m]X[k-m]\\
y(t) = & \sqrt{x(t)} \Rightarrow Y[k] = \frac{1}{2Y[0]} \bigg[ X[k] - \displaystyle \sum_{m=1}^{k-1}Y[m]Y[k-m]\bigg]
\end{array}\right\} \label{DTrules}$%
}
\end{equation}
\end{small}
Note that, the transformed functions are represented by capital letters to discriminate between the original and DT functions; and throughout this paper the similar representation is used. Moreover, the time $`t'$ symbol is also omitted for simplicity.

\section{Modeling of DFIG-based Wind Power System using Differential Transform (DT)} \label{proposedDTs}
\begin{figure}[h]
\centering
\includegraphics[width=0.45\textwidth]{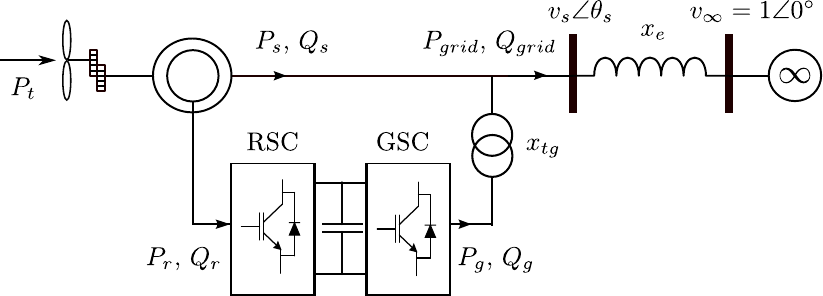}
\caption{Grid connected DFIG-based wind turbine generator}
\label{SMIBfig}
\end{figure}

Consider a grid connected single machine infinite bus system as shown in Figure \ref{SMIBfig}. The mathematical models of converters, electrical and mechanical components are required to capture the realistic response during time-domain simulation. The considered DFIG-based wind power system consists of a turbine, drive train, induction generator and the back-to-back converter system. The modeling of these components is well established in the literature \cite{salman,akhmatov2003analysis,hu2015modeling,ymishra,fmei}, but to cover every aspect the mathematical modeling of these components is discussed in the subsequent sections.

\subsection{Turbine and Drive Train Model}
The time domain differential equations of two mass drive train model are taken from \cite{ymishra, salman, fmei} and given in \eqref{turbineeq1}-\eqref{turbineeq3} (see Appendix \ref{A}). The derived differential transformed functions (DTs) of \eqref{turbineeq1}-\eqref{turbineeq3} using \eqref{DTrules} can be expressed as follows:

\begin{small}
\begin{IEEEeqnarray}{rCl}
\omega_{R}[k+1] & = & \frac{1}{2H_{g}(k+1)}\Big [ K_{sh}\Theta_{TW}[k] + C_{sh} \omega_{el} \big(\omega_{T}[k] \nonumber \\ && - \omega_{R}[k] \big)  - T_{E}[k]  \Big] \\
\omega_{T}[k+1] & = & \frac{1}{2H_{t}(k+1)}\Big [ T_{M}[k] - K_{sh}\Theta_{TW}[k] \nonumber \\ && - C_{sh}\omega_{el} \big (\omega_{T}[k] - \omega_{R}[k] \big)  \Big]\\
\Theta_{TW}[k+1] & = & \frac{\omega_{el}}{(k+1)}\Big [ \omega_{T}[k] - \omega_{R}[k]  \Big] 
\end{IEEEeqnarray}
\end{small}
The DTs of algebraic equations expressed in \eqref{Te_eq}, \eqref{Tm_eq}, \eqref{Pt_eq} and \eqref{Cppu_eq} (see Appendix \ref{A}) can be expressed as given in \eqref{DT_Te_eq}, \eqref{DT_Tm_eq}, \eqref{DT_Pt_eq} and \eqref{DT_Cppu_eq}, respectively.
\begin{small}
\begin{IEEEeqnarray}{rCl} \label{DT_Te_eq}
&& T_{E}[k] = \sum_{m=0}^{k} \frac{\big\{E_{QS}^{'}[m]I_{QS}[k-m]  + E_{DS}^{'}[m]I_{DS}[k-m] \big\}}{\omega_{s}} \\ \label{DT_Tm_eq}
&& T_{M}[k]  =  \frac{1}{\omega_{T}[0]} \bigg [P_{T}[k] - \sum_{m=0}^{k-1}T_{M}[m]\omega_{T}[k-m]\bigg]\\ \label{DT_Pt_eq}
&& P_{T}[k]  =  k_{opt}\bigg (\frac{v_{w}}{v_{wB}}\bigg )^{3}C_{PPU}[k] \\  \label{DT_Cppu_eq}
&& C_{PPU}[k] -  B_{2}[0]B_{4}[k]-B_{4}[0]B_{2}[k]
= \sum_{m=1}^{k-1}B_{2}[m]B_{4}[k-m] \nonumber\\ && +\frac{1.3801875}{v_{w}}\omega_{R}[k]
\end{IEEEeqnarray}
\end{small}
where
\begin{small}
\begin{IEEEeqnarray}{rCl}
&& B_{4}[k] - B_{4}[0]B_{3}[k] = \sum_{m=1}^{k-1} \bigg(\frac{k-m}{k}\bigg) B_{4}[m]B_{3}[k-m] \nonumber \\
&& B_{3}[k] + \frac{280}{1299} v_{w}B_{1}[k] = 0.735\varrho[k] \nonumber \\
&& B_{2}[k] - 1.283927808v_{w}B_{1}[k] = -9.7697\varrho[k] \nonumber \\ 
&& B_{1}[k] = \frac{1}{\omega_{R}[0]}\bigg[\varrho[k] - \sum_{m=0}^{k-1}B_{1}[m]\omega_{R}[k-m]\bigg] \nonumber
\end{IEEEeqnarray}
\end{small}
\subsection{Induction Generator Model}
As the dynamics of DFIG is of interest, so the single machine infinite bus system is considered for better understanding. The dynamics of the considered test system can be represented by the DAEs in \textit{p.u.} on machine base \cite{ymishra, fmei} and can be expressed as given in \eqref{gen1}-\eqref{gen6} (see Appendix \ref{A}). Equations \eqref{gen1}-\eqref{gen4} and \eqref{gen5}-\eqref{gen6} are the differential and algebraic equations, respectively. The algebraic equations represent the interface with the power system network or interface with the infinite bus. The numerical values of all the parameters are taken from \cite{ymishra, fmei}. The DTs of \eqref{gen1}-\eqref{gen4} are derived using \eqref{DTrules} and can be expressed as presented in \eqref{DTgen1}-\eqref{DTgen4}. 
\begin{small}
\begin{IEEEeqnarray}{rCl} \label{DTgen1}
&& I_{QS}[k+1] =  \frac{\omega_{el}}{(k+1)L_{s}^{'}}\Bigg [ \omega_{s}L_{s}^{'}I_{DS}[k] -R_{1}I_{QS}[k]   - \frac{1}{\tau_{r}\omega_{s}}E_{DS}^{'}[k] \nonumber \\ && + \frac{1}{\omega_{s}} \sum_{m=0}^{k}\omega_{R}[m]E_{QS}^{'}[k-m] - V_{QS}[k]  + K_{mrr}V_{QR}[k] \Bigg] \\ \label{DTgen2}
&& I_{DS}[k+1]  =   \frac{\omega_{el}}{(k+1)L_{s}^{'}}\Bigg [ -R_{1}I_{DS}[k] - \omega_{s}L_{s}^{'}I_{QS}[k]  - V_{DS}[k] \nonumber \\ &&+ \frac{1}{\omega_{s}} \sum_{m=0}^{k}\omega_{R}[m]E_{DS}^{'}[k-m] 
+ \frac{1}{\tau_{r}\omega_{s}}E_{QS}^{'}[k] + K_{mrr}V_{DR}[k] \Bigg]
\end{IEEEeqnarray}
\end{small}
\begin{small}
\begin{IEEEeqnarray}{rCl} \label{DTgen3}
&& E_{QS}^{'}[k+1]  =  \frac{\omega_{el} \omega_{s}}{(k+1)}\Bigg [ R_{2}I_{DS}[k] - \frac{1}{\tau_{r}\omega_{s}}E_{QS}^{'}[k] + E_{DS}^{'}[k] \nonumber \\ && - \frac{1}{\omega_{s}}\sum_{m=0}^{k}\omega_{R}[m]E_{DS}^{'}[k-m]  - K_{mrr}V_{DR}[k] \Bigg]  \\ \label{DTgen4}
&& E_{DS}^{'}[k+1]   =  \frac{\omega_{el} \omega_{s}}{(k+1)}\Bigg [ - R_{2}I_{QS}[k] - \frac{1}{\tau_{r}\omega_{s}}E_{DS}^{'}[k] - E_{QS}^{'}[k] \nonumber \\ &&+  \frac{1}{\omega_{s}}\sum_{m=0}^{k}\omega_{R}[m]E_{QS}^{'}[k-m]  + K_{mrr}V_{QR}[k] \Bigg]
\end{IEEEeqnarray}
\end{small}
The DTs of algebraic equations \eqref{gen5} and \eqref{gen6} can be expressed as given in \eqref{infinitebus_eq1} and \eqref{infinitebus_eq2}, respectively.
\begin{small}
\begin{IEEEeqnarray}{rCl} \label{infinitebus_eq1}
&& V_{DS}[k]V_{Q\infty} - x_{e}P_{GRID}[k] = 0 \\  \label{infinitebus_eq2}
&& V_{QS}[k] (2V_{QS}[0] - V_{Q\infty}) - x_{e}Q_{GRID}[k] + 2V_{DS}[0]V_{DS}[k] \nonumber\\ &&
= - \sum_{m=1}^{k-1} \bigg\{ V_{QS}[m]V_{QS}[k-m] + V_{DS}[m]V_{DS}[k-m] \bigg\}
\end{IEEEeqnarray}
\end{small}

\subsection{Converter Model}
The converter has two inverters connected back-to-back through a common dc-link. These inverters are commonly known as rotor side converter (RSC) and grid side converter (GSC). The transformed function of the active power balance equation for these converters (see \eqref{activepowerbalance} in Appendix \ref{A}) can be obtained as given in \eqref{DTdclink}.
\begin{small}
\begin{equation} \label{DTdclink}
V_{DC}[k+1] = \frac{FF[k]}{C_{DC}V_{DC}[0](k+1)}  
\end{equation}
\end{small}
where, $FF[k]$ is as follows:
\begin{small}
\begin{IEEEeqnarray}{rCl}
&& FF[k] = - \sum_{m=0}^{k-1} FF[m]V_{DC}[k-m] + \sum_{m=0}^{k}\big\{ V_{DR}[m]I_{DR}[k-m]   
+\nonumber \\ &&   V_{QR}[m]I_{QR}[k-m] - V_{DG}[m]I_{DG}[k-m] -  V_{QG}[m]I_{QG}[k-m]\big\} \nonumber 
\end{IEEEeqnarray}
\end{small}
\subsection{Controllers}
This section demonstrates the two controllers namely GSC and RSC which are required to control the back-to-back converters used in DFIG system. The concept of controllers can be expressed by DAEs \cite{ymishra} as given in \eqref{controller1}-\eqref{controller13} (see Appendix \ref{A}), where, $e_{ig}^{'}=|e_{ig}^{'}|\angle \delta_{ig} = e_{ds}^{'} + je_{qs}^{'}$ is the internally generated voltage in the stator. In this work, $v_{dcref}$ is set to constant value independent of wind speed and $Q_{gref}=0$ to reduce the power rating of GSC. This implies that the reactive power can only be transmitted through the stator of DFIG and the GSC only exchange active power with the grid.
The transformed functions of \eqref{controller1}-\eqref{controller5} using DT transformation rules (\textit{i.e.} using \eqref{DTrules}) can be derived as follows:
\begin{small}
\begin{IEEEeqnarray}{rCl}
	U_{1}[k+1] & = & \frac{1}{(k+1)} \big [P_{REF}[k] - P_{GRID}[k] \big ] \\
	U_{2}[k+1] & = & \frac{1}{(k+1)} \big [\delta_{IGREF} \varrho[k] - \delta_{IG}[k] + K_{P1} (P_{REF}[k] \nonumber \\ && - P_{GRID}[k]) + K_{I1}U_{1}[k] \big ] \\
	U_{3}[k+1] & = & \frac{1}{(k+1)} \big [V_{SREF} \varrho[k] - V_{S}[k] \big ] \\
	U_{4}[k+1] & = & \frac{1}{(k+1)} \big [E_{IGREF}^{'} \varrho[k] - E_{IG}^{'}[k] + K_{I3}U_{3}[k]\nonumber \\ && + K_{P3} (V_{SREF}\varrho[k]  - V_{S}[k])  \big ] \\
	U_{5}[k+1] & = & \frac{1}{(k+1)} \big [V_{DCREF} \varrho[k] - V_{DC}[k] \big ] \\
	U_{6}[k+1] & = & \frac{1}{(k+1)} \big [Q_{GREF} \varrho[k] - Q_{G}[k] \big ]
\end{IEEEeqnarray}
\end{small}
The DTs of algebraic equations given in \eqref{controller12}-\eqref{last} can be derived using transformation rules and can be expressed as given in \eqref{DT_vqg}-\eqref{lastDT}, respectively.
\begin{small}
\begin{equation} 
\resizebox{0.45\textwidth}{!}
{%
$V_{DG}[k] = - x_{tg}(K_{P5}V_{DCREF}\varrho[k] - K_{P5}V_{DC}[k] +  K_{I5}U_{5}[k])$%
}  
\label{DT_vqg}
\end{equation}
\end{small}
\begin{small}
\begin{equation}\label{DT_vdg}
\begin{split}
V_{QG}[k] +& x_{tg}K_{P6}Q_{G}[k] - V_{S}[k]  = - V_{SREF}\varrho[k] \\ & + x_{tg}(K_{P6}Q_{GREF}\varrho[k] 
 + K_{I6}U_{6}[k]) 
\end{split}
\end{equation}
\end{small}
\begin{small}
\begin{equation}\label{DT_refpower}
P_{REF}[k] = k_{opt} \sum_{m=0}^{m_{1}} \sum_{m=m_{1}}^{k}\omega_{R}[m]\omega_{R}[m_{1}-m]\omega_{R}[k-m_{1}]
\end{equation}
\end{small}
\begin{small}
\begin{equation}\label{DT_pgrid}
\begin{split}
& P_{GRID}[k] - I_{DS}[0]V_{DS}[k] - I_{QS}[0]V_{QS}[k]  - I_{DG}[0]V_{DG}[k] \\
& - V_{DG}[0]I_{DG}[k] - I_{QG}[0]V_{QG}[k] - V_{QG}[0]I_{QG}[k]  \\ 
& = \sum_{m=0}^{k-1} \bigg \{ V_{DS}[m]I_{DS}[k-m] + V_{QS}[m]I_{QS}[k-m] \bigg \} \\
& +\sum_{m=1}^{k-1} \bigg \{ V_{DG}[m]I_{DG}[k-m] + V_{QG}[m]I_{QG}[k-m] \bigg \}  
\end{split}
\end{equation}
\end{small}
\begin{small}
\begin{equation} \label{DT_qgrid}
\begin{split}
&Q_{GRID}[k] - I_{QS}[0]V_{DS}[k] + I_{DS}[0]V_{QS}[k]  - I_{QG}[0]V_{DG}[k]\\
&- V_{DG}[0]I_{QG}[k] + I_{DG}[0]V_{QG}[k] + V_{QG}[0]I_{DG}[k] \\
&=  \sum_{m=0}^{k-1} \bigg \{ V_{DS}[m]I_{QS}[k-m] - V_{QS}[m]I_{DS}[k-m] \bigg \} \\
&+ \sum_{m=1}^{k-1} \bigg \{ V_{DG}[m]I_{QG}[k-m] - V_{QG}[m]I_{DG}[k-m] \bigg \}
\end{split}
\end{equation}
\end{small}
\begin{small}
\begin{IEEEeqnarray}{rCl} \label{DT_idg}
&& I_{DG}[k]x_{tg} + V_{QG}[k] - V_{QS}[k] = 0 \\ \label{DT_iqg}
&& I_{QG}[k]x_{tg} - V_{DG}[k] + V_{DS}[k] = 0 \\ \label{DT_idr}
&& I_{DR}[k] = \frac{E_{QS}^{'}[k]}{\omega_{s}L_{m}} - K_{mrr}I_{DS}[k] \\\label{DT_iqr}
&& I_{QR}[k] = \frac{- E_{DS}^{'}[k]}{\omega_{s}L_{m}} - K_{mrr}I_{QS}[k] 
\end{IEEEeqnarray}
\end{small}
\begin{small}
\begin{equation} 
\begin{split}
V_{R}[k] + K_{P4}E_{IG}^{'}[k] + K_{P4}K_{P3}V_{S}[k]  =  K_{P4}E_{IGREF}^{'}\varrho[k] \\ 
+ K_{P4}K_{P3}V_{SREF}\varrho[k]+ K_{P4}K_{I3}U_{3}[k] + K_{I4}U_{4}[k] 
\end{split}
\label{DT_vr}
\end{equation}
\end{small}
\begin{small}
\begin{equation} 
\begin{split}
& \delta_{R}[k] + K_{P2}\delta_{IG}[k] - K_{P2}K_{P1}(P_{REF}[k] -P_{GRID}[k])
\\ & = K_{P2}\delta_{IGREF}\varrho[k] + K_{P2}K_{I1}U_{1}[k] + K_{I2}U_{2}[k]
\end{split}
\label{DT_deltar}
\end{equation}
\end{small}
\begin{small}
\begin{equation}
\resizebox{0.45\textwidth}{!}
{%
$\begin{split}
& V_{S}[k]  - \frac{V_{QS}[0]}{V_{S}[0]}V_{QS}[k] - \frac{V_{DS}[0]}{V_{S}[0]}V_{DS}[k]
= \frac{1}{2V_{S}[0]}\bigg [ \sum_{m=1}^{k-1} \bigg \{ \\ &V_{QS}[m]V_{QS}[k-m] + V_{DS}[m]V_{DS}[k-m]
- V_{S}[m]V_{S}[k-m] \bigg \} \bigg]
\end{split}$%
}
\end{equation}
\end{small}
\begin{small}
\begin{equation}
\begin{split}
 E_{IG}^{'}[k]  =& \frac{1}{2E_{IG}^{'}[0]}\bigg [ -\sum_{m=1}^{k-1} E_{IG}^{'}[m]E_{IG}^{'}[k-m]  +  \sum_{m=0}^{k} \\ & \bigg \{ E_{QS}^{'}[m]E_{QS}^{'}[k-m]   + E_{DS}^{'}[m]E_{DS}^{'}[k-m] \bigg \} \bigg] 
\end{split}
\end{equation}
\end{small}
\begin{small}
\begin{IEEEeqnarray}{rCl}
&&\phi_{R}[k] - \psi_{R}[0]\delta_{R}[k] = \sum_{m=1}^{k-1} \bigg(\frac{k-m}{k}\bigg) \psi_{R}[m]\delta_{R}[k-m] \\
&&\psi_{R}[k] + \phi_{R}[0]\delta_{R}[k] = - \sum_{m=1}^{k-1} \bigg(\frac{k-m}{k}\bigg) \phi_{R}[m]\delta_{R}[k-m]\\
&&\phi_{IG}[k] - \psi_{IG}[0]\delta_{IG}[k] = \sum_{m=1}^{k-1} \bigg(\frac{k-m}{k}\bigg)\psi_{IG}[m]\delta_{IG}[k-m] \\
&& \delta_{IG}[k] + \frac{\psi_{IG}[k]}{\phi_{IG}[0]}  = - \sum_{m=1}^{k-1} \bigg(\frac{k-m}{k}\bigg) \frac{\phi_{IG}[m]\delta_{IG}[k-m]}{\phi_{IG}[0]}
\end{IEEEeqnarray}
\end{small}
\begin{small}
\begin{equation}
\begin{split}
\psi_{IG}[k] + \frac{\psi_{IG}[0]}{E_{IG}^{'}[0]}E_{IG}^{'}[k] & = \frac{1}{E_{IG}^{'}[0]} \bigg [ E_{QS}^{'}[k] \\&  -  \sum_{m=1}^{k-1}\psi_{IG}[m]E_{IG}^{'}[k-m] \bigg ]
\end{split}
\end{equation}
\end{small}
\begin{small}
\begin{equation} 
V_{DR}[k] - V_{R}[0]\phi_{R}[k] - \phi_{R}[0]V_{R}[k] = \sum_{m=1}^{k-1}V_{R}[m]\phi_{R}[k-m]
\end{equation}
\end{small}
\begin{small}
\begin{equation} 
V_{QR}[k] - V_{R}[0]\psi_{R}[k] - \psi_{R}[0]V_{R}[k]  =  \sum_{m=1}^{k-1}V_{R}[m]\psi_{R}[k-m]   
\end{equation}
\end{small}
\begin{small}
\begin{equation} \label{lastDT}
\begin{split}
 Q_{G}[k] &- V_{DG}[0]I_{QG}[k] - I_{QG}[0]V_{DG}[k] + V_{QG}[0]I_{DG}[k] \\ & + I_{DG}[0]V_{QG}[k]  
= \sum_{m=1}^{k-1} \bigg \{ V_{DG}[m]I_{QG}[k-m] \\ & - V_{QG}[m]I_{DG}[k-m] \bigg \} 
\end{split}
\end{equation}
\end{small}

\section{Simulation Results and Discussions} \label{resultsandcon}
In this section, the derived DTs of DFIG are tested by considering a DFIG connected to the grid through a transmission line. To validate the accuracy of the proposed approach, the proposed model is investigated through various disturbances. Further, the numerical results have been compared with traditional numerical method namely Runge-Kutta (RK-4) method. All the simulations have been performed in MATLAB environment on Intel(R) Core(TM) i5-1135G7 CPU 4.20 GHz processor with 8-GB RAM.

\subsection{Investigating Various Disturbances}
For the single machine connected to infinite bus (SMIB) DFIG system, various disturbances, \textit{viz.}, variable wind speed, variable grid reactance  and network disturbance are considered for investigation purpose and simulated for $200$ sec.
\begin{figure}[!b]
	\centering
	\includegraphics[width=0.4\textwidth]{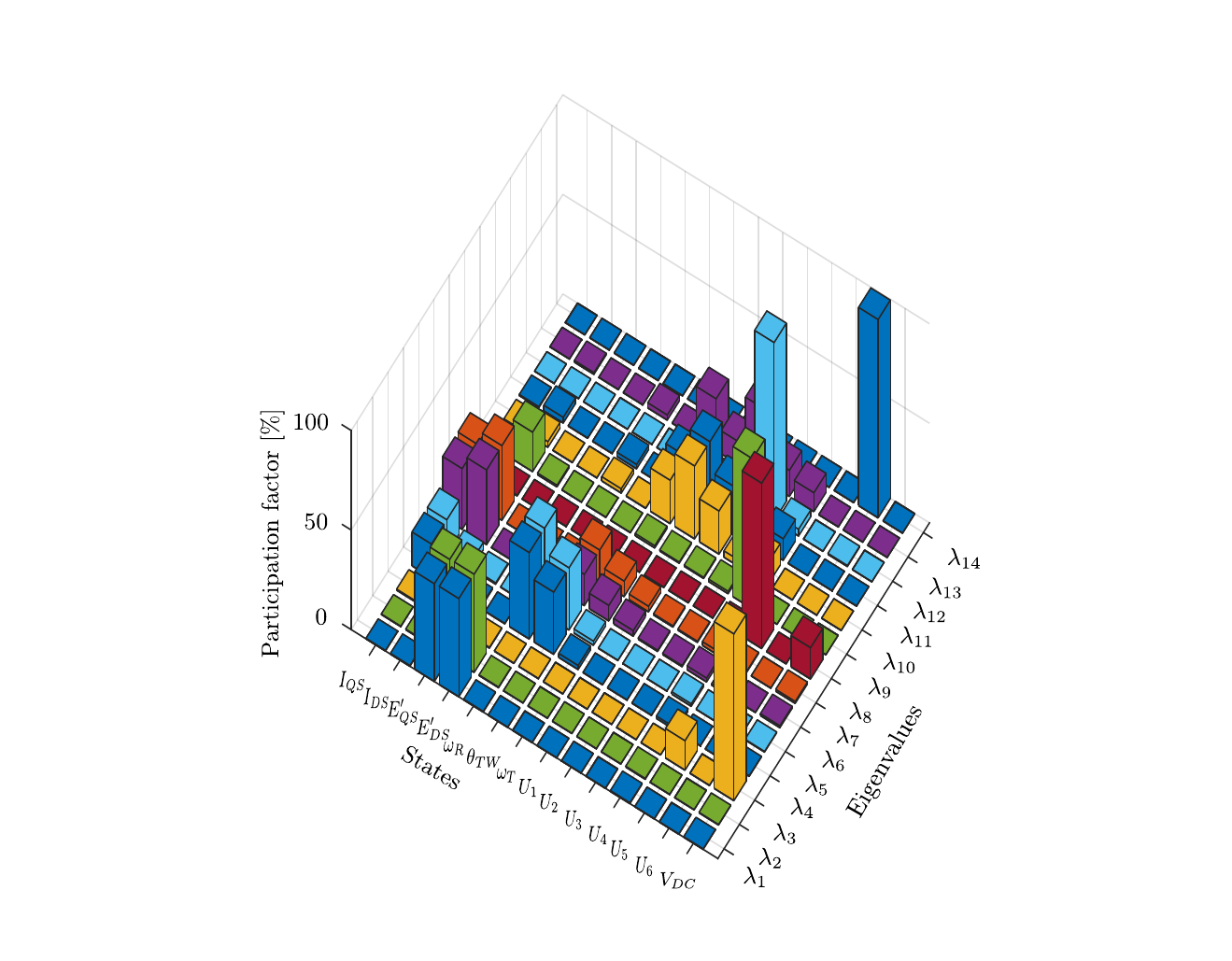}
	\caption{Participation factors of eigenvalues}
	\label{pf}
\end{figure}
\begin{figure}[!t]
	\centering
	\includegraphics{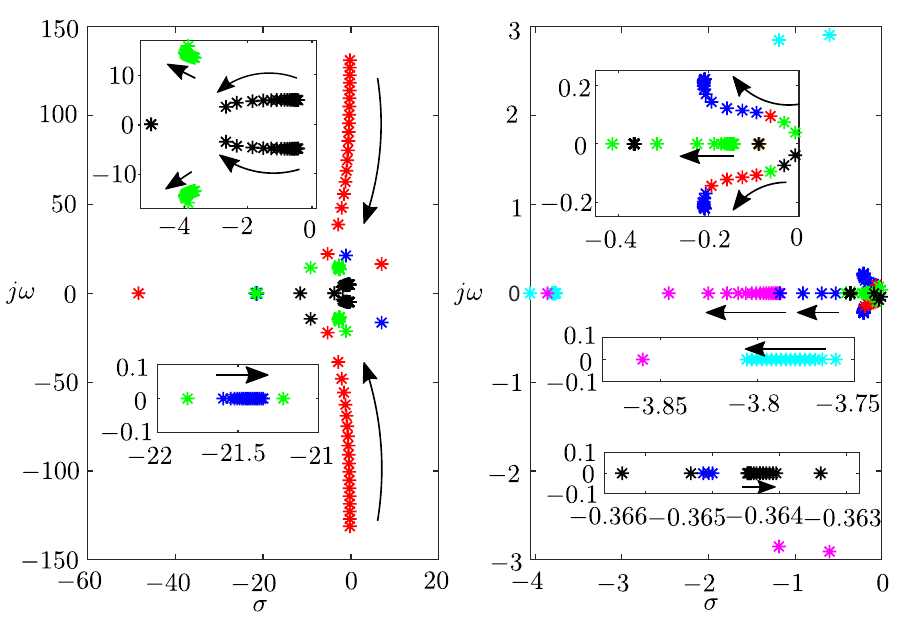}
	\caption{Eigenvalues at different wind speeds (varies from 10 to 12 m/sec.)}
	\label{Eigen}
\end{figure}

\subsubsection{Case 1: Performance under variable wind speed}
The constant wind speed is not the real case scenario as the wind speed depends on various environmental factors which are not controllable. As the wind speed is variable quantity in nature, so the performance of proposed approach should be capable to deal with the uncertain nature of wind speed. Therefore, for the investigation purpose, the system is subjected to a small perturbation by changing the wind speed at different time instants. The wind speed is considered to be $10$ \textit{m/s} from $0-2$ \textit{sec.}, $11$ \textit{m/sec.} from $2-100$ \textit{sec.}, and $10$ \textit{m/sec.} from $100-200$ \textit{sec}. 

The dynamic behavior of the DFIG-based wind power system is studied by observing the eigenvalues which are computed numerically using state matrix. The participation factors are calculated at wind speed $10$ \textit{m/sec.} and shown in Figure \ref{pf}. The movement of eigen values with the increase in wind speed is shown in Figure \ref{Eigen}. The participation factor analysis is carried out to relate the contribution of individual physical state to a particular mode. The participation factor represents the nature of modes: the $\lambda_{1}$ and $\lambda_{2}$ are the electrical modes (as electrical states $e_{qs}^{'}$, $e_{ds}^{'}$, participate in this), $\lambda_{4}$-$\lambda_{7}$ and $\lambda_{10}$-$\lambda_{11}$ are the electromechanical modes (as the electrical states $i_{qs}$ and mechanical states ($\omega_{r}$, $\omega_{t}$, $\theta_{tw}$ participate in this); and $\lambda_{3}$, $\lambda_{8}$, $\lambda_{9}$ $\lambda_{12}$, $\lambda_{13}$ and $\lambda_{14}$ are the non-oscillating modes. The electrical mode has lowest damping ratio and highest frequency among all the modes. 
\begin{figure}[!t]
    \centering
    \includegraphics{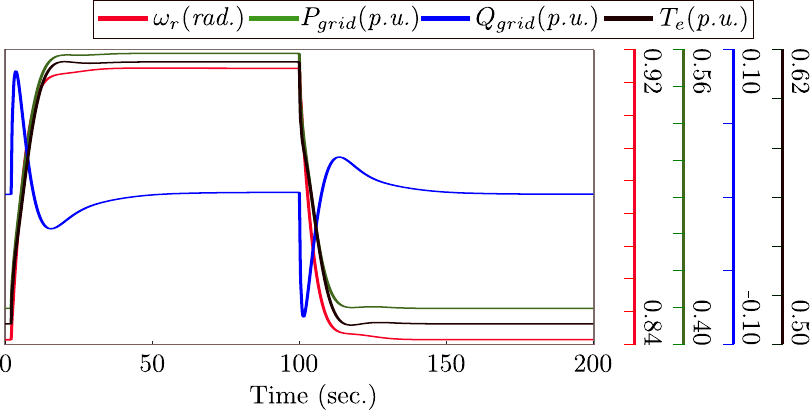}
    \caption{Dynamic response of the generator speed, electromagnetic torque, active and reactive power to the grid under variable wind speed using DT}
    \label{fig1}
\end{figure}
\begin{figure}[!t]
	\centering
	\includegraphics[width=0.45\textwidth]{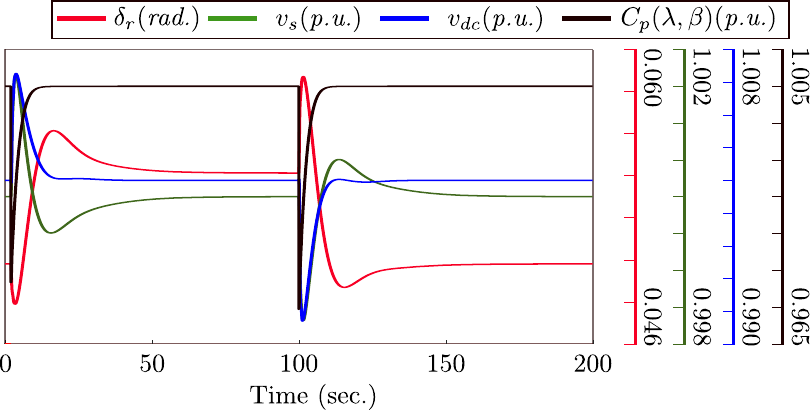}
	\caption{Dynamic response of the rotor angle, terminal voltage, dc capacitor voltage and performance coefficient under variable wind speed using DT}
	\label{fig2}
\end{figure}

The dynamic response of the various parameters (generator speed, active and reactive power delivered to the grid, electromagnetic torque, rotor angle, terminal voltage, dc-link voltage and performance coefficient of wind turbine) of the DFIG system using proposed approach are shown in Figure \ref{fig1} and \ref{fig2}. From Figure \ref{fig1} and \ref{fig2}, it can be observed that the generator speed, rotor angle, active power delivered to the grid and electromagnetic torque settled at $0.91$ \textit{p.u.}, $0.054$ \textit{rad.}, $0.56$ \textit{p.u.} and $0.61$ \textit{p.u.} respectively, as the wind speed changes to $11$ \textit{m/sec.} at 2 \textit{sec.} These parameters again settled back to previous values as the wind speed again changes to $10$ \textit{m/sec.} at $100$ \textit{sec.} The reactive power delivered to the grid and terminal voltage returns to their target values after each disturbance due to the controller action. The level of variation in dc capacitor voltage decides the rating of the converters for safe and reliable operation of the DFIG system. So, the dynamic response of the dc capacitor is also studied and presented in Figure \ref{fig2}. From this figure, it can be clearly observed that oscillations are not high and settle down very quickly. Figure \ref{fig2} also presents the dynamic response of wind turbine performance coefficient and it can be observed that the action of controllers bringing back the DFIG operation to its maximum power point tracking level after every changed operating conditions \textit{i.e.} wind speed.
\begin{figure}[!t]
	\centering
	\includegraphics[width=0.45\textwidth]{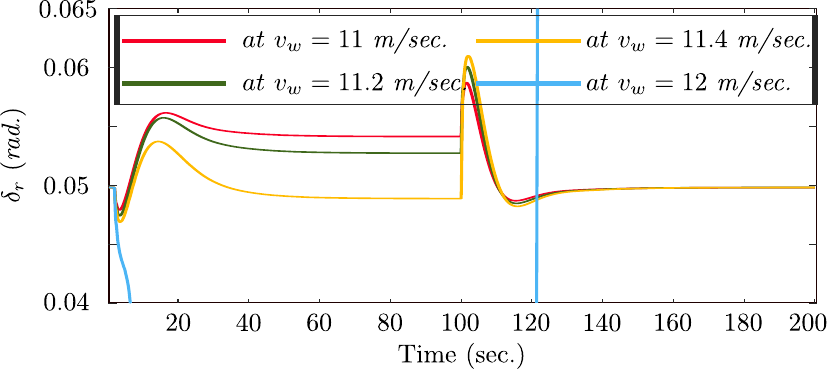}
	\caption{Dynamic response of rotor angle under different wind speeds using DT}
	\label{Windvar}
\end{figure}
\begin{figure}[!t]
	\centering
	\includegraphics[width=0.45\textwidth]{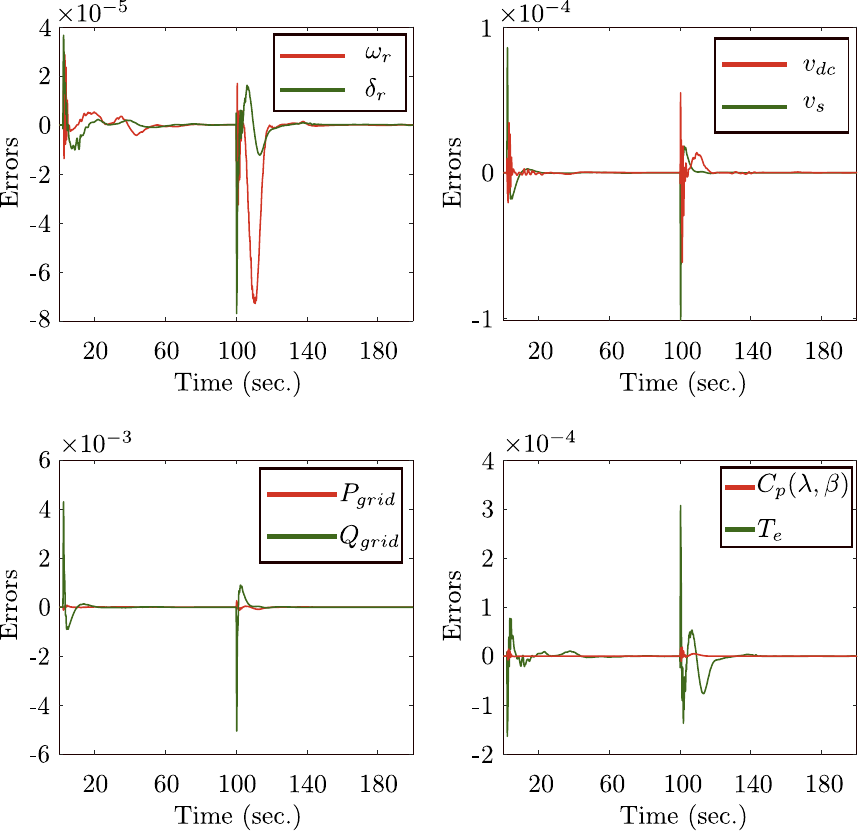}
	\caption{Maximum error of state and algebraic variables}
	\label{fig9}
\end{figure}

Further, from Figure \ref{Eigen}, it can be observed that as the wind speed increases some poles start moving right hand side and remaining poles to the left hand side. At $12$ \textit{m/sec.} wind speed, the system poles shifted to the right half plane \textit{i.e.} unstable. It can be observed that as the wind speed increases both the electrical and mechanical damping decrease, thereby increasing the oscillations. The similar conclusion can be drawn from the time-domain simulation results using DTM as shown in Figure \ref{Windvar}.

To investigate the accuracy of the proposed approach, the results obtained using DT and RK-4 method are also compared and shown in Figure \ref{fig9}. From the figure, it can be observed that the results of the proposed approach with $k=8$ and the time step length $0.01$ \textit{sec.} accurately match with the results of the RK-4 method for all the parameters. The maximum error between the values of parameters evaluated using the proposed approach and RK-4 method are $4\times10^{-5}$,  $1\times10^{-4}$, $6\times10^{-3}$ and $4\times10^{-4}$ over the $200$ \textit{sec.} simulation period. Therefore, it can be concluded that the proposed model can assess the dynamic performance of the DFIG-based wind power system accurately considering the variability of wind speed.

\subsubsection{Case 2: Performance under line outage}
\begin{figure}[!t]
	\centering
	\includegraphics[width=0.45\textwidth]{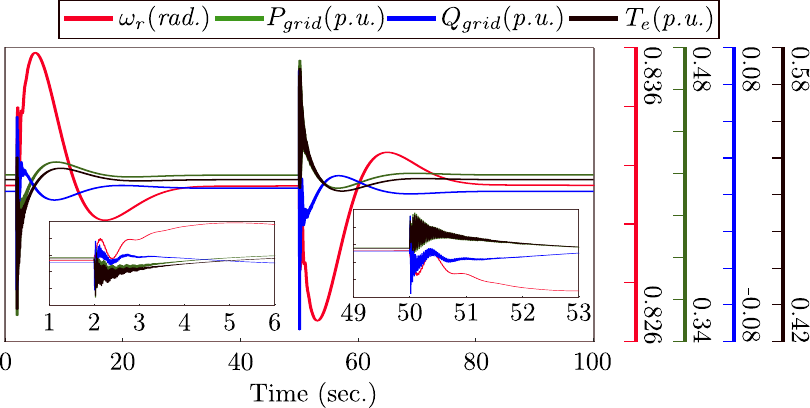}
	\caption{Dynamic response of the generator speed, electromagnetic torque, active and reactive power to the grid under variable grid reactance using DT}
	\label{fig11}
\end{figure}

\begin{figure}[!t]
	\centering
	\includegraphics[width=0.45\textwidth]{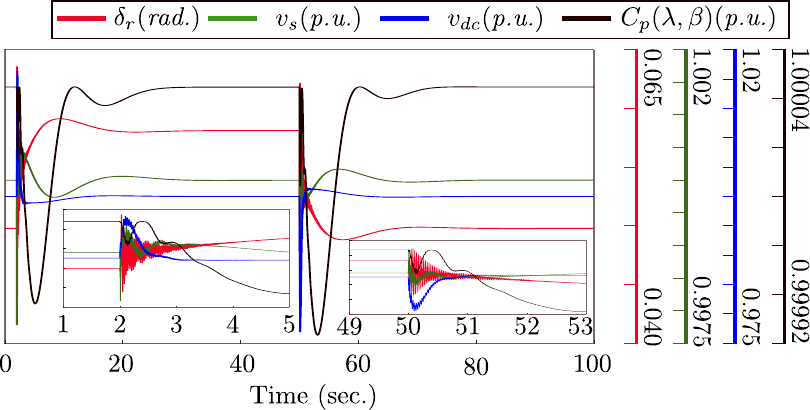}
	\caption{Dynamic response of the rotor angle, terminal voltage, dc capacitor voltage and performance coefficient under variable grid reactance using DT}
	\label{fig12}
\end{figure}

In this section, the performance of the proposed model is investigated during the variation in grid reactance \textit{i.e.} line outage. For this purpose, the value of grid reactance $x_{e}$ is changed from $0.02$ to $0.04$ \textit{p.u}. The value of grid reactance is considered to be $0.02$ \textit{p.u.} from $0-2$ \textit{sec.}, $0.04$ \textit{p.u.} from $2-50$ \textit{sec.} (due to outage of one of the grid line) and $0.02$ \textit{p.u.} from $50-100$ \textit{sec.} The dynamic response of various parameters are presented in Figures \ref{fig11} and \ref{fig12}; and the similar inferences can be drawn from these.

The eigenvalues at $10$ \textit{m/sec.} wind speed for the  grid reactance $0.02$ \textit{p.u.} and $0.04$ \textit{p.u.} has been calculated and it is observed that as the value of grid reactance is increased (\textit{i.e.} weak grid) the stability of the system is  decreased. Figure \ref{fig12} also depicts the same as the grid reactance is increased, the rotor angle is set to higher values. It can also be observed that when the grid is strong, \textit{i.e.}, $x_{e}$ is small, the oscillations are less. So, it can be concluded that the time-domain simulations results using DTM are consistent with eigenvalue analysis. 


\subsubsection{Case 3: Performance under grid disturbance}
\begin{figure}[!t]
	\centering
	\includegraphics[width=0.45\textwidth]{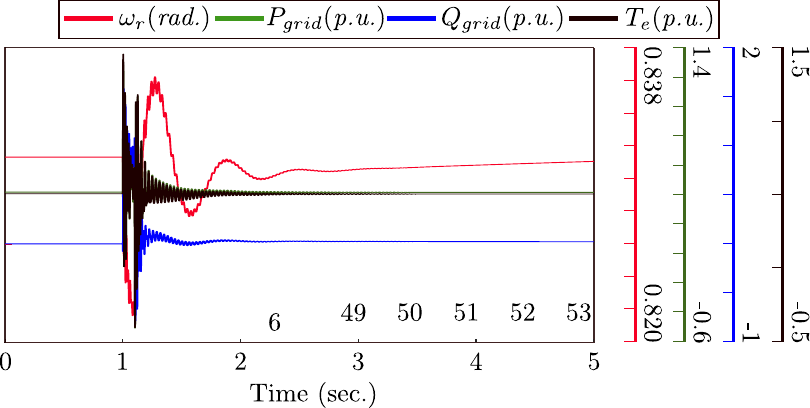}
	\caption{Dynamic response of the generator speed, electromagnetic torque, active and reactive power to the grid under grid disturbance using DT}
	\label{fig21}
\end{figure}

\begin{figure}[!t]
	\centering
	\includegraphics[width=0.45\textwidth]{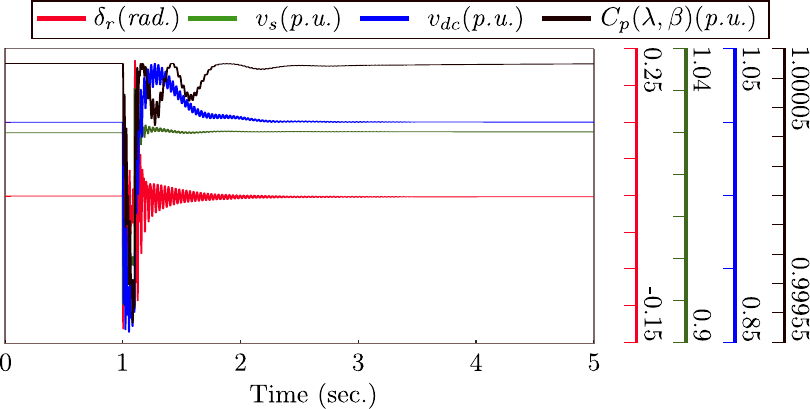}
	\caption{Dynamic response of the rotor angle, terminal voltage, dc capacitor voltage and performance coefficient under grid disturbance using DT}
	\label{fig22}
\end{figure}

The voltage of the system and the electromagnetic torque developed by the machines reduce whenever there is a fault in the network. The reduction in the system voltage and electromagnetic torque depends on various factors such as location of the fault, type of fault, fault impedance \textit{etc.} The imbalance between electromagnetic torque developed and the input mechanical torque leads to either acceleration or deceleration of the rotor. The time domain analysis has been carried out to recommend remedial action before the system loose synchronism. Therefore, the performance of the proposed approach is also investigated with grid disturbance. For this purpose, it is assumed that due to the grid fault, the voltage magnitude of infinite bus is $0.92$ \textit{p.u.} for a duration of $0.1$ \textit{sec}, after that the infinite bus voltage magnitude restores to the nominal value \textit{i.e.} $1$ \textit{p.u.}. The dynamic response of various parameters are presented in Figures \ref{fig21} and \ref{fig22}. After, $0.1$ \textit{sec.} the fault is cleared, now as per the network configuration the rotor angle settles to its new equilibrium position. The oscillations of the rotor angle and speed depend on the damping offered by the system. From these figures, it can be observed that due to the advancement of rotor during fault, the voltage phase angle are changed so the power oscillations in the system can be observed. The terminal voltage falls to $0.92$ \textit{p.u.} during the fault and restores to its nominal value $1$ \textit{p.u.} around 0.5 \textit{sec}.
 
From the simulations results, it can be concluded that the proposed model can accurately assess the dynamic response of the DFIG-based wind power system under different DFIG control and operating conditions.

\subsection{DT vs. RK-4 Method}
To investigate the numerical stability, the derived DTs have been tested with different time step length. The maximum time step length to maintain the numerical stability for both the methods has been obtained by gradually increasing the time step length until numerical instability occurs. The maximum time step length for case 1 with proposed approach and RK-4 method are $0.53$ and $0.51$ respectively, beyond these limits the respective methods will diverge. Further, the DT method requires $1.6652$ \textit{sec.} execution time and $195$ number of steps to simulate the considered system over the period of $200$ \textit{sec.} However, the RK-4 method requires $1.8573$ \textit{sec.} execution time and $231$ number of steps to simulate the same scenario. For case 2 and case 3, the maximum time step length with DT method are $0.8971$ and $0.2356$ respectively, while with RK-4 method are $0.8514$ and $0.2284$ respectively. The execution time required by DT method are $3.9854$ and $2.5506$ \textit{sec.}; and number of steps taken by the DT method are $3459$ and $1574$ for $100$ \textit{sec.} and $5$ \textit{sec.} simulation period respectively. However, the execution time required by RK-4 method are $4.2892$ and $2.5506$ \textit{sec.}; and number of steps taken by the RK-4 method are $4164$ and $1701$ for case 2 and case 3 respectively. From these simulations results, it can be observed that the maximum time step length of DT method is larger than the RK-4 method and the computation time of the DT method is reduced by around $9$\% as compared to RK-4 method. Therefore, it can be concluded that the DT method is more stable and reduces the execution time as compared to the traditional numerical RK-4 method. 

\section{Conclusion} \label{conclusions}
This paper proposed a differential transform method based approach for time-domain simulation of DFIG-based wind power system. In this research work, the differential transforms of the differential and algebraic equations involved in the DFIG-based wind power system are derived. Further, to investigate the derived DTs, the numerical simulations have been carried out for grid connected DFIG-based wind power system with various disturbances. The simulations results shows that the proposed approach increases the time step length hence reduces the execution time by maintaining the accuracy which is comparable to the traditional numerical RK-4 method. Therefore, it can be concluded that the proposed approach has great potential for fast power system simulation. The influence on the performance of the proposed model when the wind-mills are interconnected to the multi-machine power system will be our next focus of the research.


%


%
%

\ifCLASSOPTIONcaptionsoff
  \newpage
\fi



\bibliographystyle{IEEEtran}
\bibliography{IEEEabrv,REF}
\newpage
\appendices
\section{DAEs of DFIG based wind turbine generator system \cite{ymishra, salman, fmei}} \label{A}
\textbf{Turbine and Drive Train} 
\begin{small}
\begin{IEEEeqnarray}{rCl} \label{turbineeq1}
\frac{d\omega_{r}}{dt} & = & \frac{1}{2H_{g}}\Big [ T_{sh} + C_{sh} \omega_{el} (\omega_{t} - \omega_{r}) - T_{e}  \Big] \\ \label{turbineeq2}
\frac{d\omega_{t}}{dt} & = & \frac{1}{2H_{t}}\Big [ T_{m} - T_{sh} - C_{sh} \omega_{el} (\omega_{t} - \omega_{r})  \Big] \\  \label{turbineeq3}
\frac{d\theta_{tw}}{dt} & = & \omega_{el} (\omega_{t} - \omega_{r}) 
\end{IEEEeqnarray}
\end{small}
where,
\begin{small}
\begin{IEEEeqnarray}{rCl} 
&& T_{sh}  =  K_{sh}\theta_{tw} \\ \label{Te_eq}
&& T_{e}   =  (e_{qs}^{'}i_{qs}+e_{ds}^{'}i_{ds})/\omega_{s} \\ \label{Tm_eq}
&& T_{m}   =  P_{t}/\omega_{t} \\  \label{Pt_eq}
&& P_{t}   =  k_{opt}(v_{w}/v_{wB})^{3}C_{ppu}(\lambda,\beta)\\ \label{Cp_eq}
&& C_{p}(\lambda,\beta)  =  c_{10}\lambda + c_{1}\bigg(\frac{c_{2}}{\lambda+c_{8}\beta}-\frac{c_{2}c_{0}}{\beta^{3}+1}-c_{3}\beta \nonumber \\ && - c_{4}\beta^{c_{5}}-c_{6}\bigg)
exp\bigg(\frac{-c_{7}}{\lambda+c_{8}\beta}\bigg)
\end{IEEEeqnarray}
\end{small}
The value of parameters ($c_{0}-c_{10}$ \textit{etc.}) are taken from \cite{fmei} and after simplification, \eqref{Cp_eq} can be expressed in terms of state variable ($\omega_{r}$) as follows:
\begin{small}
\begin{flalign} \label{Cppu_eq}
C_{ppu}(\lambda,\beta)  = \bigg(\frac{1.283927808v_{w}}{\omega_{r}} - 9.7697\bigg) \nonumber\\
exp\bigg(\frac{-280v_{w}}{1299\omega_{r}}+0.735\bigg)
 + \frac{1.3801875\omega_{r}}{v_{w}} 
\end{flalign}
\end{small}
\textbf{Induction Generator}
\begin{small}
\begin{IEEEeqnarray}{rCl}\label{gen1}
\frac{di_{qs}}{dt} & = & \frac{\omega_{el}}{L_{s}^{'}}\Big [ -R_{1}i_{qs} + \omega_{s}L_{s}^{'}i_{ds} + \frac{\omega_{r}e_{qs}^{'}}{\omega_{s}} - \frac{e_{ds}^{'}}{\tau_{r}\omega_{s}} - v_{qs} \nonumber \\ &&+K_{mrr}v_{qr} \Big] \\  \label{gen2}
\frac{di_{ds}}{dt} & = &   \frac{\omega_{el}}{L_{s}^{'}}\Big [ -R_{1}i_{ds} - \omega_{s}L_{s}^{'}i_{qs} + \frac{\omega_{r}e_{ds}^{'}}{\omega_{s}} + \frac{e_{qs}^{'}}{\tau_{r}\omega_{s}} - v_{ds} \nonumber \\ &&+K_{mrr}v_{dr} \Big] \\\label{gen3}
\frac{de_{qs}^{'}}{dt} & = &  \omega_{el}\omega_{s}\Big [ R_{2}i_{ds} - \frac{e_{qs}^{'}}{\tau_{r}\omega_{s}} + \Big ( 1- \frac{\omega_{r}}{\omega_{s}} \Big ) e_{ds}^{'} - K_{mrr}v_{dr} \Big] \\\label{gen4}
\frac{de_{ds}^{'}}{dt} & = &  \omega_{el}\omega_{s}\Big [ K_{mrr}v_{qr} - R_{2}i_{qs} - \frac{e_{ds}^{'}}{\tau_{r}\omega_{s}} - \Big ( 1- \frac{\omega_{r}}{\omega_{s}} \Big ) e_{qs}^{'}  \Big] \\
\label{gen5}
v_{ds}&v_{q\infty}& - x_{e}P_{grid} = 0 \\ \label{gen6}
v_{qs}^{2} & + & v_{ds}^{2} - v_{qs}v_{q\infty} - x_{e}Q_{grid} =  0 
\end{IEEEeqnarray}
\end{small}
\textbf{Converter and Controllers}
\begin{small}
\begin{equation} \label{activepowerbalance}
\underbrace{v_{dr}i_{dr}+v_{qr}i_{qr}}_{P_{r}} = \underbrace{v_{dg}i_{dg}+v_{qg}i_{qg}}_{P_{g}} + \underbrace{C_{dc}v_{dc}\frac{dv_{dc}}{dt}}_{P_{dc}}
\end{equation}
\end{small}
\begin{small}
\begin{IEEEeqnarray}{rCl} \label{controller1}
\frac{du_{1}}{dt} & = & P_{ref}  - P_{grid} \\ \label{controller2}
\frac{du_{2}}{dt} & = & \delta_{igref}  - \delta_{ig} + K_{P1} (P_{ref}  - P_{grid}) + K_{I1}u_{1} \\ 
\frac{du_{3}}{dt} & = & |v_{s}|_{ref}  - |v_{s}| \\ \label{controller3}
\frac{du_{4}}{dt} & = & |e_{ig}^{'}|_{ref} - |e_{ig}^{'}| + K_{P3} (|V_{s}|_{ref} - |V_{s}|)  + K_{I3}u_{3} \IEEEeqnarraynumspace \\ \label{controller4}
\frac{du_{5}}{dt} & = & |v_{dc}|_{ref} - |v_{dc}| \\ \label{controller5}
\frac{du_{6}}{dt} & = & Q_{gref} - Q_{g} 
\end{IEEEeqnarray}
\end{small}
\begin{small}
\begin{IEEEeqnarray}{rCl} \label{controller12}
v_{dg} & = & - x_{tg} \big [ K_{P5} (|v_{dc}|_{ref} - |v_{dc}|) + K_{I5}u_{5}\big]\\ \label{controller13}
v_{qg} & = &   x_{tg} \big [ K_{P6} (Q_{gref} - Q_{g}) + K_{I6}u_{6}\big] - |v_{s}|_{ref} + |v_{s}| \IEEEeqnarraynumspace
\end{IEEEeqnarray}
\end{small}
The other relationships among different parameters are as follows:
\begin{small}
\begin{IEEEeqnarray}{rCl}\label{refpower}
P_{ref} & = & k_{opt}\omega_{r}^{3} \\ \label{controller6}
P_{grid}& = & \underbrace{v_{ds}i_{ds}+v_{qs}i_{qs}}_{P_{s}} + \underbrace{v_{dg}i_{dg}+v_{qg}i_{qg}}_{P_{g}} \\ \label{controller7}
Q_{grid} & = & \underbrace{v_{ds}i_{qs}-v_{qs}i_{ds}}_{Q_{s}} + \underbrace{v_{dg}i_{qg}-v_{qg}i_{dg}}_{Q_{g}} \\ \label{controller8}
i_{dg} & = & (v_{qs}-v_{qg})/x_{tg} \\\label{controller9}
i_{qg} & = & (v_{dg}-v_{ds})/x_{tg} \\\label{controller10}
i_{dr} & = & (e_{qs}^{'}/\omega_{s}L_{m})-K_{mrr}i_{ds}\\\label{controller11}
i_{qr} & = & -(e_{ds}^{'}/\omega_{s}L_{m})-K_{mrr}i_{qs} \\\label{controller15}
v_{r} & = & K_{P4}\big \{|e_{ig}^{'}|_{ref} - |e_{ig}^{'}| +K_{P3} (|v_{s}|_{ref} - |v_{s}| )\big \} \nonumber \\&& + K_{P4}K_{I3} u_{3} + K_{I4}u_{4} \\\label{controller14}
\delta_{r} & = & K_{P2}\big \{|\delta_{ig}|_{ref} - |\delta_{ig}| +K_{P1} (P_{ref} - P_{grid} )\big \} \nonumber \\&& + K_{P2}K_{I1} u_{1} + K_{I2}u_{2}  \\
v_{s}^{2} & = & v_{qs}^{2} + v_{ds}^{2}  \\
e_{ig}^{'2} & = & e_{qs}^{'2} + e_{ds}^{'2}  \\
\phi_{r} & = & sin\delta_{r} \\
\psi_{r} & = & cos\delta_{r} \\
\phi_{ig} & = & sin\delta_{ig} \\
\psi_{ig} & = & cos\delta_{ig} \\
\psi_{ig} & = & \frac{e^{'}_{qs}}{e^{'}_{ig}} \\
v_{dr} & = & v_{r}sin\delta_{r} \\ 
v_{qr} & = & v_{r}cos\delta_{r} \\ 
Q_{g} & = & v_{dg}i_{qg} - v_{qg}i_{dg} \label{last}
\end{IEEEeqnarray}
\end{small}

\end{document}